\documentclass[aps,prl,twocolumn,amsmath,amssymb,nofootinbib,superscriptaddress,floatfix,reprint,longbibliography]{revtex4-1}
\usepackage[dvips]{graphicx}
\usepackage{latexsym}
\usepackage{amsmath}
\usepackage{amsfonts}
\usepackage{amssymb}
\usepackage{color}
\usepackage{txfonts}
\usepackage{float}
\usepackage{url}
\usepackage[colorlinks=true, urlcolor=blue, linkcolor=blue, citecolor=blue]{hyperref}
\usepackage{ulem}
\usepackage{tikz-feynman}
\usepackage{tikz,tikz-feynhand} 
\usepackage{graphicx}
\usepackage{extpfeil}
\usepackage{subfigure}
\usepackage{physics}
\usepackage{titlesec}
\titleformat{\section}{\normalfont\normalsize\bfseries\centering}{\thesection.}{1em}{}
\begin{document}
	\newcommand{\fig}[2]{\includegraphics[width=#1]{#2}}
	\newcommand{\la}{{\langle}}
	\newcommand{\ra}{{\rangle}}
	\newcommand{\dg}{{\dagger}}
	\newcommand{\upa}{{\uparrow}}
	\newcommand{\dna}{{\downarrow}}
	\newcommand{\ab}{{\alpha\beta}}
	\newcommand{\ias}{{i\alpha\sigma}}
	\newcommand{\ibs}{{i\beta\sigma}}
	\newcommand{\hH}{\hat{H}}
	\newcommand{\hn}{\hat{n}}
	\newcommand{\hc}{{\hat{\chi}}}
	\newcommand{\hU}{{\hat{U}}}
	\newcommand{\hV}{{\hat{V}}}
	\newcommand{\br}{{\bf r}}
	\newcommand{\bR}{{\bf R}}
	\newcommand{\bA}{{\bf A}}
	\newcommand{\bk}{{{\bf k}}}
	\newcommand{\bq}{{{\bf q}}}
	\newcommand{\ri}{{\mathrm{i}}}
	\def\gsim{~\rlap{$>$}{\lower 1.0ex\hbox{$\sim$}}}
	\setlength{\unitlength}{1mm}
	\newcommand{\pprl}{Phys. Rev. Lett. \ }
	\newcommand{\pprb}{Phys. Rev. {B}}

\title {Paramagnetic contribution in superconductors with different-mass Cooper pairs}
\author{Pengfei Li}
\affiliation{Beijing National Laboratory for Condensed Matter Physics and Institute of Physics,
	Chinese Academy of Sciences, Beijing 100190, China}
\affiliation{School of Physical Sciences, University of Chinese Academy of Sciences, Beijing 100190, China}

\author{Kun Jiang}
\email{jiangkun@iphy.ac.cn}
\affiliation{Beijing National Laboratory for Condensed Matter Physics and Institute of Physics,
	Chinese Academy of Sciences, Beijing 100190, China}
\affiliation{School of Physical Sciences, University of Chinese Academy of Sciences, Beijing 100190, China}

\author{Jiangping Hu}
\email{jphu@iphy.ac.cn}
\affiliation{Beijing National Laboratory for Condensed Matter Physics and Institute of Physics,
	Chinese Academy of Sciences, Beijing 100190, China}
\affiliation{Kavli Institute of Theoretical Sciences, University of Chinese Academy of Sciences,
	Beijing, 100190, China}
 \affiliation{New Cornerstone Science Laboratory, 
	Beijing, 100190, China}

\date{\today}

\begin{abstract}
 Cooper pairs formed by two electrons with different effective mass are common in multiband superconductors, pair density wave states  and other superconducting systems with multi-degrees of freedom.
In this work, we show that there are paramagnetic contributions to the superfluid stiffness in superconductors with different-mass Cooper pairs. This paramagnetic response is owing to the relative motion between two electrons with different mass.Additionally, we identify a particle-hole dichotomy effect related to this mass difference. We investigate the paramagnetic contributions based on the linear response theory in two-band superconductors with interband pairings  and in  pair density wave states respectively. Our results offer a new perspective to the electromagnetic superfluid stiffness in unconventional superconductors beyond traditional BCS response.
\end{abstract}

\maketitle

Superconductors (SCs)  are defined by  zero resistance and Meissner effect with perfect diamagnetism \cite{schrieffer}.  
Microscopically, the central ingredients for superconductors are the Cooper pairs and their phase coherence, in which electrons bind together two by two and condense to form a coherent quantum state \cite{schrieffer,bcs_theory}. Especially, the phase coherence plays an essential role in the electromagnetic response of SCs. Any phase disturbance induced by external magnetic fields is disfavored by Cooper pairs' phase coherence leading to the diamagnetic Meissner effect.
The diamagnetic rigidity in SCs are normally characterized by the superfliud stiffness $\rho_s$, which is defined in the London equation coefficient as $\mathbf{j}=-\frac{4e^2}{\hbar^2}\rho_s\mathbf{A}=\frac{1}{\lambda^2}\mathbf{A}$ \cite{schrieffer,coleman2015introduction,shoucheng}. 
Theoretically, $\rho_s$ contains a paramagnetic current correlation function part and a diamagnetic part. And the paramagnetic part is strictly zero due to superconducting gap in the traditional BCS response \cite{schrieffer,bcs_theory}. How to go beyond this superfliud stiffness response is not only one interesting question in BCS theory, but also a window towards understanding high-temperature superconductors
\cite{xiang2022d,Uemura2,ivan_nature,invan_review,Armitage_PhysRevLett.122.027003,Hirschfeld_PhysRevResearch.2.013228,Qianghua_PhysRevLett.128.137001,zixiang}.

On the other hand, Cooper pairs are formed by two electrons. Since effective mass of electrons can vary dramatically in a solid state system, there are superconducting pairing systems with Cooper pairs formed by two electrons with different effective mass, namely the different-mass Copper pair systems. The different-mass Copper pair systems are common in superconductors. For example, in multiband systems \cite{suhl_PhysRevLett.3.552}, interband pairings between two different electron  bands generically  provide different-mass Copper pairs; a pair density wave  state (PDW) \cite{pdw} with vector $Q$ connecting $k$ and $-k+Q$ sectors can easily host different mass Cooper pairs as well.
Therefore, it is natural to ask whether new properties can be associated to the different-mass pairing.
In this work, we demonstrate that Cooper pairs with different masses exhibit an unexpected paramagnetic response. This response is directly related to the particle and hole components of the superconductor belonging to different mass sectors. Additionally, this mass difference induces a particle-hole dichotomy effect, which has been recently confirmed in monolayer FeSe SCs \cite{dichotomy}.

\begin{figure}[t]
	\begin{center}
		\fig{3.4in}{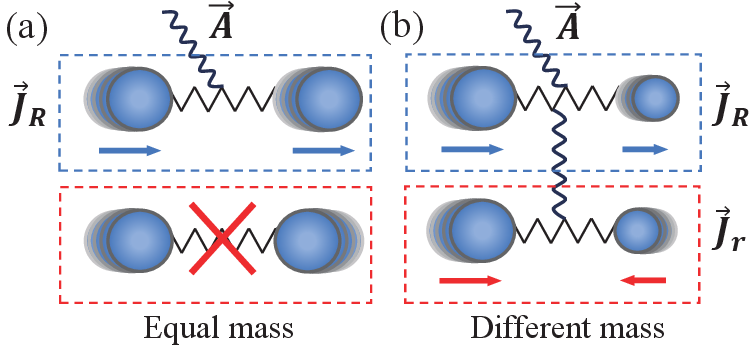}
		\caption{
		Schematic diagram of the response current to the electromagnetic field of the two-electron pair with (a) equal mass, (b) different mass. In case (a), there is no $\bf{J}_r$ contribution and gauge field $\mathbf{A}$ couples with $\bf{J}_R$. In case (b), the $\bf{J}_r$ contributes owing to its coupling with $\bf{J}_R$. $\bf{J}_r$ plays the key role in paramagnetic response.
			\label{fig1}}
	\end{center}
		\vskip-0.5cm
\end{figure}

To investigate the electromagnetic response of SC with different-mass Cooper pairs, we start from how a system with two different-mass electrons couples to the electromagnetic gauge field $\bA$ from a Cooper problem \cite{Cooper_PhysRev.104.1189}. The kinetic part Hamiltonian of this Cooper problem can be written as $H_c = \sum_i\frac{1}{2m_i}\left[-\ri\bf{\nabla}_{\br_i}-e\bA(\br_i)\right]^2$. 
Using the center-of-mass (COM) coordinate $\bR=\frac{\br_1+\br_2}{2} $, the relative coordinate $\bar{\br}= \br_1-\br_2$ and their corresponding mass $M_\pm^{-1}= m_1^{-1}\pm m_2^{-1}$,
the Hamiltonian $H_c$ transforms into
\begin{eqnarray}
    H_c&=& \frac{1}{2M_+} \left[-\ri\frac{\nabla_{\bR}}{2}-e\bA(\bR)\right]^2 + \frac{\left(-\ri\nabla_{\bar{\br}}\right)^2}{2M_-} + \nonumber\\
    &&\qquad\qquad \frac{1}{2M_-} \left\{-\ri\frac{\nabla_{\bR}}{2}-e\bA(\bR),-\ri\nabla_{\bar{\br}}\right\}.\label{energy}
\end{eqnarray}

From Eq. \ref{energy}, we can see the first term describing the energy of COM coupled to the gauge field $\mathbf{A}$, which contains both the paramagnetic and diamagnetic contribution. Note that we approximate the vector potential as $\bA(\bR\pm\bar{\br}/2) \approx \bA(\bR)$, due to the slowly varying field. The second term $\left(-\ri\nabla_{\bar{\br}}\right)^2/2M_-$ describes the kinetic energy of the relative motion without coupling to the gauge field  because the relative motion is a charge-less process. The last term $\frac{1}{2M_-} \left\{-\ri\frac{\nabla_{\bR}}{2}-e\bA(\bR),-\ri\nabla_{\bar{\br}}\right\}$ describes the coupling between the COM motion and relative motion, which plays an important role in linking the relative motion to the gauge field.

The current is obtained with two contribution from COM and relative motions as $\mathbf{J} = -\partial H_c /\partial \bA=\bf{J}_R+\bf{J}_r$, as schematically plotted in Fig.\ref{fig1}.
As shown in Fig.\ref{fig1}(a), there only remains the current $\bf{J}_R$ from COM motion when $M_-=0$. 
The electromagnetic response of COM shows perfect diamagnetism, where the paramagnetic response is strictly zero.
This is the standard conclusion from the traditional BCS theory.
On the other hand, for the Cooper pair with different mass plotted in Fig.\ref{fig1}(b), both $\bf{J}_R$ in the up panel and $\bf{J}_r$ in the lower panel exist. COM part performs like the Cooper pair composed of the equal mass electrons.  However, the relative motion contributes a paramagnetic response, resulting in the reduction of the superfluid stiffness owing to the vanishing diamagnetic term related to $M_-$ in $\left(-\ri\nabla_{\bar{\br}}\right)^2/2M_-$. 



\begin{figure}
	\begin{center}
		\fig{3.4in}{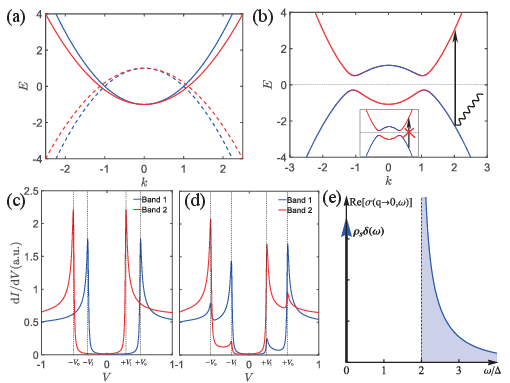}
		\caption{(a) The energy of the two parabolic bands with different mass. The dashed line is the particle-hole symmetric band. (b) The band structure of the two different-mass band superconductor with inter-band pairing. The arrow represents the direct transition between the two asymmetric Bogoliubov band due to the optical absorption. Inset: the case of the two bands with the equal mass where the optical process is forbidden. (c) The projected DOS onto the two bands of interband pairing model. (d) The projected DOS considering the intraband pairing $\Delta_0=0.1$. (e) The complex optical conductance spectrum of the system. The zero frequency peak $\rho_s\delta(\omega)$ is from superfluid while the area under $\sigma(\omega\neq0)$ is proportional to paramagnetic contribution $\Pi_{\mu\nu}$. The parameters used in the calculation are $\gamma_1=1$, $\gamma_2=0.8$, $\mu=1$ and $\Delta_1=0.4$. 
			\label{fig2}}
	\end{center}
		\vskip-0.5cm
\end{figure}

To illustrate above ideas, we first study the response of multiband SCs with interband pairing. The multiband effect in superconductivity is a long-term topic \cite{suhl_PhysRevLett.3.552}, since the discovery of BCS theory. The multiple band signatures have been widely observed in superconductors, including elemental metals Nb, Ta, V and Pb \cite{Nb_Ta_V,Pb}, MgB$_2$ \cite{mgb2}, doped-SrTiO$_3$ \cite{STO}, especially the iron pnictides and chalcogenides \cite{iron1,iron2,iron_review}. 
Recently, we also find the interband pairing in monolayer FeSe \cite{dichotomy}.
Here, we consider a simplified two-band model with interband pairing:
\begin{equation}
H_{\mathbf{k}}=\sum_{\alpha, \sigma} \xi_\alpha(\mathbf{k}) c_{\mathbf{k}, \alpha, \sigma}^{\dagger} c_{\mathbf{k}, \alpha, \sigma}+\Delta_1 \sum_{\alpha \neq \beta}\left(c_{\mathbf{k}, \alpha, \uparrow}^{\dagger} c_{-\mathbf{k}, \beta, \downarrow}^{\dagger}+\text { H.c. }\right) \label{model}
\end{equation}
where $\alpha,\beta=1,2$ label two separated bands. These two bands host the dispersion $\xi_\alpha(k)=\frac{k^2}{2m_\alpha}-\mu$ with mass $m_\alpha$ and chemical potential $\mu$, as the red and blue parabolic bands illustrated in Fig.\ref{fig2}(a).
$\Delta_1$ is the mean-field interband pairing order parameter. Normally, this interband pairing should not be the leading pairing instability. 
Since we focus on the responses rather than the mechanism, we take a phenomenology approach to this problem and assume the interband pairing is dominated in this toy model.
Below we define $\gamma_\alpha=\frac{1}{2m_\alpha}$ for convenience.  Under the basis $\Psi_k^\dagger=\{c_{1k\uparrow}^\dagger,c_{2k\uparrow}^\dagger,c_{1,-k\downarrow},c_{2,-k\downarrow}\}^\mathrm{T}$, the BdG Hamiltonian can be written as
\begin{equation}
H_{BdG}=\left(\begin{array}{cccc}
\gamma_1 k^2-\mu & 0 & 0 & \Delta_1 \\
0 & \gamma_2 k^2-\mu & \Delta_1 & 0 \\
0 & \Delta_1 & -\gamma_1 k^2+\mu & 0 \\
\Delta_1 & 0 & 0 & -\gamma_2 k^2+\mu
\end{array}\right). \label{HBdG}
\end{equation} 
where $k^2=k_x^2+k_y^2$. The Hamiltonian is composed of two decoupled blocks. 
Thus, we can focus on the $H_{1}$ block for convenience as 
\begin{equation}
H_{1}=\left(\begin{array}{cc}
\gamma_1 k^2-\mu  & \Delta_1 \\
\Delta_1  & -\gamma_2 k^2+\mu
\end{array}\right). \label{Hhalf}
\end{equation} 

The eigenvalue of $H_1$ are $E_k^\pm=\eta_k\pm \sqrt{\epsilon_k^2+\Delta_1^2}$,
where $\eta_k=\frac{\gamma_1-\gamma_2}{2}k^2$ and $\epsilon_k=\frac{\gamma_1+\gamma_2}{2}k^2-\mu$.
Note that the system is fully gapped only when $\Delta_1>\frac{\mu\abs{\gamma_1-\gamma_2}}{2\sqrt{\gamma_1\gamma_2}}$, namely $E_k^+>0$ and $E_k^-<0$. Fig.\ref{fig2}(b) shows the band dispersion of $H_1$ block in the fully-gapped region. The color of the band represents the weight of the two band in Fig.\ref{fig2}(a).It's obvious that the two band is not particle-hole symmetric because the particle and hole parts of Eq.\ref{Hhalf} come from different mass sector with the relative motion $\eta_k$ contribution.

Interestingly, this  relative motion can lead to a particle-hole dichotomy effect in the tunneling DOS as shown in Fig.\ref{fig2}(c). The system exhibits four coherence peaks located at $\pm V_\mathrm{o}$ and $\pm V_\mathrm{i}$. The blue and red colors represent the projected tunneling DOS onto bands 1 and 2, respectively, which behave very differently. For band 1, the coherence peaks are located at positions $-V_\mathrm{i}$ and $+V_\mathrm{o}$, with no quasiparticle weight at positions $-V_\mathrm{o}$ and $+V_\mathrm{i}$. Conversely, for band 2, the coherence peaks are located at positions $-V_\mathrm{o}$ and $+V_\mathrm{i}$. Notice that, only $\Delta_1$ is one limiting case. Generally speaking, the intraband pairing $\Delta_0$ always exists. Including this fact, the projected DOS for both bands exhibit all the four coherence peaks as shown in Fig.\ref{fig2}(d), where the particle-hole dichotomy effect persists. For the inner coherence peaks, band 1 exhibits a higher peak at $-V_\mathrm{i}$, while band 2 shows a higher peak at $+V_\mathrm{i}$. For the outer coherence peaks, the situation is completely reversed. 
This particle-hole dichotomy effect is deeply linked to the recent experimental observed sublattice dichotomy effect in the monolayer two-gap FeSe SC \cite{dichotomy}. For example, the coherence peak of $\alpha$-Fe sublattice at $-V_\mathrm{i}$ is apparent larger than $V_\mathrm{i}$ while the $\beta$-Fe sublattice shows a reversed feature.



Next, we want to find the superfluid stiffness. Within linear response theory, the response current to the vector potential of electromagnetic field can be obtained via $J_\mu(\bq,\Omega)=-\sum_{\nu}K_{\mu\nu}(\bq,\Omega)A_\nu(\bq,\Omega)$. Here $K_{\mu\nu}$ is the electromagnetic response tensor with two part of contributions, paramagnetic response from current-current correlation function $\Pi_{\mu \nu}$ and diamagnetic response from charge density $\langle\frac{\hat{n}}{m}\rangle$ as $K_{\mu\nu}=-e^2(\Pi_{\mu \nu}+\langle\frac{\hat{n}}{m}\rangle)$.  According to London equation,the superfluid stiffness can be defined via $\rho_s=\frac{\hbar^2}{4e^2}\Re K_{\mu\mu}(\omega=0,q\xrightarrow{}0)$.
Notice that the paramagnetic current obtained from Eq.\ref{Hhalf} can be decomposed into two components as discussed above: the COM motion and the relative motion,
\begin{eqnarray}
J^P(q)&=&J^{P}_R+J^{P}_r=
\frac{\gamma_1+\gamma_2}{2} (2k+q) \sigma_0 + \frac{\gamma_1-\gamma_2} {2}(2k+q) \sigma_3 \nonumber
\end{eqnarray}
Then the current-current correlation function can be calculated by $\Pi_{\mu\nu}(q,\ri \Omega_n) = \frac{1}{V\beta}\sum_{k,\ri\omega_n} \Tr[J^P_\mu \mathcal{G} J^P_\nu \mathcal{G}]$, where $\mathcal{G}$ is the Green's function for $H_1$. More details of calculations can be found in Supplemental Material (SM) \cite{SM}.

At zero temperature, considering the complete model given in Eq.\ref{HBdG}, the paramagnetic contribution of the superfluid response of this system with interband pairing is
\begin{equation}
    \Re \Pi_{\mu\nu}(\Omega=0,q\to 0) = -\frac{2(\gamma_1-\gamma_2)^2 \epsilon_F N_{\gamma^+}(0)}{\gamma_1+\gamma_2}\delta_{\mu\nu} \label{para}
\end{equation}
where $N_{\gamma^+}$ is the DOS of the energy dispersion $\epsilon_k$, which is proportional to $\frac{1}{\gamma_1+\gamma_2}$. 
This nonzero paramagnetic contribution is the key finding in this work. We can also calculate the diamagnetic contribution $\langle\frac{\hat{n}}{m}\rangle=2(\gamma_1+\gamma_2)n_c$, where $n_c$ is the number of carriers in either band. 

The paramagnetic contrition $\Re \Pi_{\mu\nu}(\Omega=0,q\to 0)$ vanishes if $\gamma_1=\gamma_2$, which corresponds to the Cooper pairs composed of the equal mass electrons.
This is consistent with the conventional BCS theory, where the $\Pi_{\mu \nu}$ contribution is strictly zero owing to the SC gap as discussed above \cite{schrieffer}.
Hence, the SC shows perfect diamagnetism from $\langle\frac{\hat{n}}{m}\rangle$ due to the vanishing $\Pi_{\mu \nu}$.
On the other hand, if $\gamma_1 \neq \gamma_2$, the paramagnetic response $\Pi_{\mu \nu}$ remains finite at zero temperature. Although the whole SC system remains diamagnetism, the existence of relative motion current $J^{P}_r(q)$ reduces the superfluid stiffness with finite $\Pi_{\mu \nu}$.

Actually the above results can be understood from the optical perspective. In the optical response, the optical conductivity $\sigma(\omega)$ follow a sum rule $\int_{0}^{\infty}\Re\sigma(\omega)=\langle \frac{\pi e^2}{2m}\hat{n}\rangle$ \cite{FGT1,FGT2}. For clean SCs in BCS theory, all the optical weight transfers into the $\omega=0$ with $\rho_s\delta(\omega)$ with $\Re\sigma(\omega\neq0)=0$. This $\Re\sigma(\omega\neq0)=0$ is due to the forbidden optical selection rule from particle-hole symmetry, as proved in Ref. \cite{optical_nagaosa} and illustrated in Fig.\ref{fig2}(b) inset. However, this situation changes in the multiband system \cite{optical_nagaosa}.
The $\sigma(\omega\neq0)$ becomes finite, as calculated in Fig.\ref{fig2}(d). To satisfy optical sum rule, the superfluid stiffness in $\delta(\omega)$ must have finite $\Pi_{\mu \nu}$ contribution. 

\begin{figure}
	\begin{center}
		\fig{3.4in}{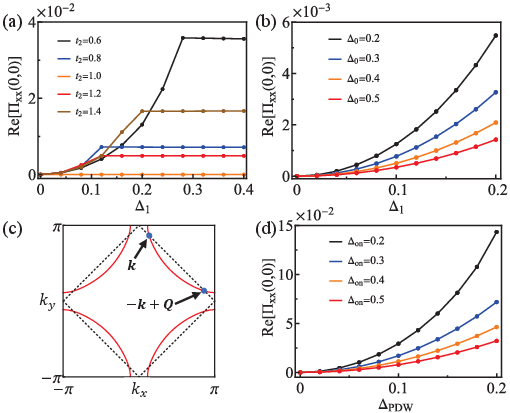}
		\caption{(a) The paramagnetic response of the two-band lattice model with only interband pairing for different hopping parameter. (b) The paramagnetic response of the two-band lattice model with both intraband and interband pairing. (c) The first Brillouin zone (BZ) of the square lattice. The red lines represent the Fermi surface (FS) of the dispersion in Eq.\ref{PDW}. The arrows show the $k$ on the FS and its counterpart $-k+Q$ forming PDW Cooper pairs with $Q=(\pi,\pi)$. The dashed lines are the reduced BZ. (d) The paramagnetic response of the $(\pi,\pi)$-PDW with finite onsite pairing.
			\label{fig3}}
	\end{center}
		\vskip-0.5cm
\end{figure}

To further demonstrate above analytical results, we can numerically calculate the paramagnetic response using lattice models. Here we consider two band model on square lattice with the energy dispersion $\xi_{1,2}=-2t_{1,2}\left(\cos{k_x}+\cos{k_y}\right)-\mu$, where $t_{1,2}$ is the nearest-neighbor hopping parameter for band 1,2 respectively. We set $t_1=1$ as an energy scale. The ratio of the effective mass, defined by $m^*=\left(\frac{\partial^2 \xi_k}{\partial k^2}\right)^{-1}$ of two energy bands is $m^*_1/m^*_2=t_2/t_1$.
The paramagnetic responses $\Re\Pi_{xx}(\Omega=0,q\to 0)$ as a function of interband pairing order parameter $\Delta_1$ for different value of $t_2$ are plotted in in Fig.\ref{fig3}(a). Notice that, as we fix intraband pairing to zero, the system is gapless with Bogoliubov Fermi surface contributed by the lower quasiparticle band when $\Delta_1$ is small. Here we only focus on the interband transition process as shown by the arrow in Fig.\ref{fig2}(b) corresponding to the analytical result in Eq.\ref{para}, although the Bogoliubov Fermi surface also contributes to the paramagnetic response through the intraband process. (See SM \cite{SM} for details of numerical calculations and discussions of intraband transition process due to the Bogoliubov Fermi surface.)

From Fig.\ref{fig3}(a), we can find that the paramagnetic response is always zero for any $\Delta_1$ at $t_1=t_2$, which corresponds to $m_1=m_2$. For $t_1 \neq t_2$, the interband process starts to generate finite response. 
For small value of interband pairing, the $\Pi_{xx}$ response strengthens as $\Delta_1$ increases. 
This is because the only the quasiparticles under the Fermi surface contribute to the paramagnetic response, whose number  increases as $\Delta_1$ increases. After $\Delta_1$ exceeds a critical value $\Delta_c$, the Bogoliubov Fermi surface disappears, and all quasiparticles in the lower band participate in interband processes, leading to a saturated response. The evolution of the energy spectrum with $\Delta_1$ is shown in SM \cite{SM}.
We can also find that the saturated response is positively correlated with $\abs{t_1-t_2}$ and negatively correlated with $t_1+t_2$. These are consistent with the our analytical calculation results based on the effective continuum model.

In more realistic multiband SC cases such as iron based SCs, the intraband pairing is always the leading instability, which always dominates in comparison to the interband pairing \cite{irongroup}. The influence of the intraband pairing to $\Pi_{\mu \nu}$ becomes important.
We plot the paramagnetic response for different intraband pairing order parameter $\Delta_0$ in Fig.\ref{fig3}(b) with fixed hopping parameter $t_1=1$ and $t_2=0.6$. The system is fully gapped for all parameters in the calculation. So the response completely results from interband processes. The result in Fig.\ref{fig3}(b) suggests that the interband pairing strengthens the paramagnetic response, while the intraband pairing suppresses it as $\Delta_0$ increasing from $0.2$ to $0.5$.
Thus, although the paramagnetic response is small in the intraband pairing $\Delta_0$ dominated multiband system, it does exist as long as the interband pairing is finite.

Pair density wave is another important example for different-mass Copper pairs \cite{pdw}. Recently, the PDW has been widely explored in CsV$_3$Sb$_5$, cuprates and other superconducting system \cite{pdw_135,pdw_vortex,pdw_2212,pdw_yayu,pdw_lee}.
PDW is a special SC state composed of the Copper pairs with momentum $k$ and $-k\pm Q$. The effective mass of band electrons are naturally different for electrons in PDW Cooper pairs due to the finite momentum $Q$.
To simplify our discussion, we calculate the paramagnetic response in a PDW with $Q=(\pi,\pi)$ on square lattice. 
As the pure PDW state may host a Bogoliubov Fermi surface, we also add an onsite pairing term to achieve a gap system. The Hamiltonian $H_{PDW}$ under the basis $\Psi_k^\dagger=\{c_{k\uparrow}^\dagger,c_{k+Q\uparrow}^\dagger,c_{-k\downarrow},c_{-k+Q\downarrow}\}^\mathrm{T}$ can be written as
\begin{equation}
H_{PDW}=\left(\begin{array}{cccc}
\xi_k & 0 & \Delta_{on} & \Delta_{PDW} \\
0 & \xi_{k+Q} & \Delta_{PDW} & \Delta_{on} \\
\Delta_{on} & \Delta_{PDW} & -\xi_{-k} & 0 \\
\Delta_{PDW} & \Delta_{on} & 0 & -\xi_{-k+Q}
\end{array}\right). \label{PDW}
\end{equation}
where $\xi_k=-2t\left(\cos{k_x}+\cos{k_y}\right)-4t'\cos{k_x}\cos{k_y}-\mu$. In the calculation we set $t=1$, $t'=-0.3$ and $\mu=-1$.  Fig.\ref{fig3}(c) shows the Fermi surface of $\xi_k$ and the reduced Brillouin zone (BZ). $\Delta_{on}$ and $\Delta_{PDW}$ represents the onsite pairing order parameter and the PDW order parameter respectively. 
The paramagnetic responses as a function of $\Delta_{PDW}$ for different $\Delta_{on}$ are shown in Fig.\ref{fig3}(d). The result is similar to the case in Fig.\ref{fig3}(b). As $\Delta_{PDW}$ dominating different-mass pairing increases, the $\Pi_{\mu \nu}$ keep increasing as the interband cases. On the other hand, the onsite pairing term suppresses the paramagnetic response because it leads to the pairing of electrons with opposite momenta, i.e., electrons with the equal mass.

\begin{figure}
	\begin{center}
		\fig{3.4in}{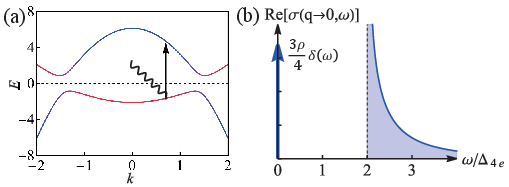}
		\caption{(a) The single particle excitation of the charge $4e$ SC model using the wavefunction method \cite{charge4e}. The arrow represents the direct transition between the two bands due to the optical absorption. (b) The complex optical conductance spectrum of the system. The parameters used in the calculation are $m=0.5$, $\mu=2$ and $\Delta_{4e}=1$.
			\label{fig4}}
	\end{center}
		\vskip-0.5cm
\end{figure}

Above results can be further extended to charge $4e$ superconductors \cite{charge4e}, where the cooper pairs are formed by four electrons. Notice that the one electron excitation here is equivalent to a three-hole excitation rather than a one-hole excitation in charge $2e$ SCs. 
Using the wavefunction method \cite{charge4e}, we can calculate the Green's function of charge $4e$ SC as
\begin{eqnarray}
    G^R(k\sigma,\omega)_{\alpha\alpha}&=&\frac{u_k^2}{\omega+\mathrm{i}\eta-(E_k-\xi_k)}+\frac{v_k^2}{\omega+\mathrm{i}\eta+(E_k+\xi_k)}, \nonumber
\end{eqnarray}
where $\xi_k=\frac{k^2}{2m}-\mu$ and $E_k=\sqrt{4\xi_k^2+\Delta_{4e}^2}$. The single particle excitation spectrum is shown in Fig.\ref{fig4}(a). This spectrum is highly particle-hole asymmetric, since adding one electron and adding one hole are related to different excitation states \cite{charge4e}. 
From linear response, we find the superfluid density reduces to $3/4$ of that in a charge 2e SCs.
Through the optical conductance shown in Fig.\ref{fig4}(b), we found that finite paramagnetic response corresponding to the direct transition between the two excitations at high frequency leads to a reduction in the superfluid density. Although it stems from the multi-body pairing rather than different-mass Cooper pairs, the reduction of the superfluid stiffness due to finite paramagnetic contribution is a common property in unconventional superconductors.

In summary, we carry out a systematic study of electromagnetic property in superconductors with different-mass Cooper pairs. We find that the different-mass Cooper pairing can result in new paramagnetic contributions in superfluid stiffness. This paramagnetic response  directly links to the relative motion between two different mass electrons inside each Cooper pair. Using the two-band model with interband pairing, the paramagnetic responses are calculated based on the linear-response theory from both continuum model and lattice model. This paramagnetic response is finite only at $m_1 \neq m_2$. Additionally, this mass difference induces a particle-hole dichotomy effect in the tunneling DOS. Furthermore, this reduced superfluid stiffness can be understood from the optical sum rule. 
Besides the interband pairing in multiband systems, this paramagnetic response also exists in PDW system and other different-mass Cooper pair SCs.
Finally, we want to point out this paramagnetic contribution is common both in different-mass Cooper pair systems and  multi-body pairing superconducting systems \cite{charge4e,CBCS}. These results offer a new perspective on the superfluid stiffness in superconductors beyond traditional BCS response.

\subsection{Acknowledgement}
This work is supported by the Ministry of Science and Technology  (Grant No. 2022YFA1403901), the National Natural Science Foundation of China (Grant No. NSFC-12174428), the Strategic Priority Research Program of the Chinese Academy of Sciences (Grant No. XDB28000000, XDB33000000), the New Cornerstone Investigator Program, and the Chinese Academy of Sciences through the Project for Young Scientists in Basic Research (2022YSBR-048).

\bibliography{reference}

\clearpage
\onecolumngrid
\begin{center}
	\textbf{\large Supplemental Material}
\end{center}

\setcounter{equation}{0}
\setcounter{figure}{0}
\setcounter{table}{0}
\setcounter{page}{1}
\makeatletter
\renewcommand{\theequation}{S\arabic{equation}}
\renewcommand{\thefigure}{S\arabic{figure}}
\renewcommand{\bibnumfmt}[1]{[S#1]}
\newcommand{\tE}{\tilde{E}}
\newcommand{\tu}{\tilde{u}}
\newcommand{\tv}{\tilde{v}}

\section{Superfluid response and optical conductance}
The eigenenergy of the Hamiltonian $H_1$ (Eq.5) in the main text is
\begin{equation}
	E_k^\pm=\eta_k\pm \mathcal{E}_k=\eta_k\pm \sqrt{\epsilon_k^2+\Delta^2}
\end{equation}
with corresponding eigenvectors $(u_k,v_k)^\mathrm{T}$ and $(v_k,-u_k)^\mathrm{T}$ respectively in which
\begin{eqnarray}
	u_k&=&\sqrt{\frac{1}{2}\left(1+\frac{\epsilon_k}{\mathcal{E}_k}\right)} \\
	v_k&=&\sqrt{\frac{1}{2}\left(1-\frac{\epsilon_k}{\mathcal{E}_k}\right)}
\end{eqnarray}
The Green's function of the system described by $H_1$ can be expressed as
\begin{equation}
	\mathcal{G}=\left(\ri\omega_n-H_1\right)^{-1} \equiv\left(\begin{array}{cc}
		G(k,\ri\omega_n) & F(k,\ri\omega_n) \\
		F^\dagger(k,\ri\omega_n) & \bar{G}(k,\ri\omega_n)
	\end{array}\right)
\end{equation}
where 
\begin{eqnarray}
	G(k,\ri\omega_n)&=&\frac{u_k^2}{\ri\omega_n-E_+}+\frac{v_k^2}{\ri\omega_n-E_-}, \\
	\bar{G}(k,\ri\omega_n)&=&\frac{v_k^2}{\ri\omega_n-E_+}+\frac{u_k^2}{\ri\omega_n-E_-}, \\ 
	F(k,\ri\omega_n)&=&\frac{u_k v_k}{\ri\omega_n-E_+}-\frac{u_k v_k}{\ri\omega_n-E_-}.
\end{eqnarray}
The paramagnetic current is defined as
\begin{equation}
	\hat{J}^P_\mu(q)=\sum_{\alpha,k} \gamma_\alpha(2k+q)_\mu c_{k+q,\alpha}^\dagger c_{k\alpha}
\end{equation}
which can be decomposed into two components as discussed in the main text: the COM motion and the relative motion,
\begin{eqnarray}
	J^P(k,q)&=&J^{P}_R+J^{P}_r=
	\frac{\gamma_1+\gamma_2}{2} (2k+q) \sigma_0 + \frac{\gamma_1-\gamma_2} {2}(2k+q) \sigma_3 
\end{eqnarray}

Then the paramagnetic part of the linear response, i.e. current-current correlation function, can be calculated by 
\begin{eqnarray}
	\Pi_{\mu\nu}(q,\ri \Omega_n)&=& \frac{1}{V\beta}\sum_{k,\ri\omega_n} \Tr[J^P_\mu(q)\mathcal{G}(k,\ri\omega_n)J^P_\nu(q)\mathcal{G}(k+q,\ri\omega_n+\ri\Omega_n)] \label{JJ} \\
	&=& (\gamma_1+\gamma_2)^2 \chi_{00} + 2(\gamma_1^2-\gamma_2^2) \chi_{30} + (\gamma_1-\gamma_2)^2 \chi_{33}
\end{eqnarray}
where
\begin{equation}
	\chi_{ij}(q,\ri\Omega_n)=\frac{1}{4V\beta}\sum_{k,\ri\omega_n} (2k+q)_\mu(2k+q)_\nu \Tr \left[\sigma_i \mathcal{G}(k,\ri\omega_n)\sigma_j \mathcal{G}(k+q,\ri\omega_n+\ri\Omega_n)\right].
\end{equation}
Considering the case $T=0$, the results of these correlation functions are
\begin{eqnarray}
	\chi_{00}(q,\ri \Omega_n)&=& \frac{1}{4V}\sum_{k}(2k+q)_\mu(2k+q)_\nu \left[-\frac{u^2\tv^2+v^2\tu^2-2uv\tu\tv }{\ri\Omega_n+E^+-\tE^-}  +\frac{\tu^2v^2+u^2\tv^2-2uv\tu\tv}{\ri\Omega_n+E^--\tE^+} \right].
\end{eqnarray}
\begin{eqnarray}
	\chi_{33}(q,\ri \Omega_n)&=& \frac{1}{4V}\sum_{k}(2k+q)_\mu(2k+q)_\nu \left[-\frac{u^2\tv^2+v^2\tu^2+2uv\tu\tv }{\ri\Omega_n+E^+-\tE^-}  +\frac{\tu^2v^2+u^2\tv^2+2uv\tu\tv}{\ri\Omega_n+E^--\tE^+} \right].
\end{eqnarray}
\begin{eqnarray}
	\chi_{30}(q,\ri \Omega_n)&=& \frac{1}{4V}\sum_{k}(2k+q)_\mu(2k+q)_\nu \left[-\frac{u^2\tv^2-v^2\tu^2}{\ri\Omega_n+E^+-\tE^-}  +\frac{\tu^2v^2-u^2\tv^2}{\ri\Omega_n+E^--\tE^+} \right].
\end{eqnarray}
where $u=u_k$, $v=v_k$, $E^\pm=E_k^\pm$, $\tu=u_{k+q}$, $\tv=v_{k+q}$ and $\tE^\pm=E^\pm_{k+q}$.

In the long wavelength limit $q \to 0$, we have $\tu \to u$, $\tv\to v$ and $\tE^\pm \to E^\pm$ and obviously $\chi_{00}=\chi_{30}=0$ and then
\begin{eqnarray}
	\Re \Pi_{\mu\nu}(\ri\Omega_n=0,q\to 0) &=& -\frac{4}{V}\sum_{k}k_\mu k_\nu \frac{(\gamma_1-\gamma_2)^2u^2v^2}{\mathcal{E}_k} \nonumber\\
	&=& -\frac{(\gamma_1-\gamma_2)^2}{V}\sum_{k}k_\mu^2 \frac{\Delta^2}{\mathcal{E}_k^3} \delta_{\mu\nu} \nonumber\\
	&=& -\frac{(\gamma_1-\gamma_2)^2 }{2(\gamma_1+\gamma_2)} \int d\epsilon N_{\gamma^+}(\epsilon)\epsilon_F \frac{\Delta^2}{(\epsilon^2+\Delta^2)^{3/2}} \delta_{\mu\nu} \nonumber\\
	&=& -\frac{(\gamma_1-\gamma_2)^2 \epsilon_F N_{\gamma^+}(0)}{\gamma_1+\gamma_2}\delta_{\mu\nu} \label{para}
\end{eqnarray}
where $N_{\gamma_1+\gamma_2}$ is the DOS of the 2D free electron gas with energy dispersion $(\gamma_1+\gamma_2)k^2$, which is proportional to $\frac{1}{\gamma_1+\gamma_2}$. If $\gamma_1=\gamma_2$, the paramagnetic contrition vanishes which is just the single band case. This result suggests that the nonzero paramagnetic contribution stems from interband pairing.

Then we shall consider the diamagnetic contribution. The diamagnetic current is defined as
\begin{equation}
	J^D_\mu=\sum_{k\alpha}\frac{e^2 \hat{n}_{\alpha, \mu\nu}}{m_\alpha}A_\nu
\end{equation}
The matrix form is
\begin{equation}
	J^D(q)=\left(\begin{array}{cc}
		2\gamma_1  & 0 \\
		0  & -2\gamma_2
	\end{array}\right)
	= J^{D}_R+J^{D}_r
	= (\gamma_1+\gamma_2)\sigma_3+(\gamma_1-\gamma_2)\sigma_0
\end{equation}
Thus the diamagnetic contribution of superfluid stiffness can be obtained by calculating
\begin{eqnarray}
	\left<\bf{J}^D\right>&=&\frac{1}{V\beta}\sum_{k,\ri\omega_n} \Tr \left[J^D \mathcal{G}(k,\ri\omega_n)\right] \bf{A} \nonumber\\
	&=& \sum_k(\gamma_1+\gamma_2)\left[1-\frac{\epsilon_k}{\mathcal{E}_k}\left[n_F(E_k^-)-n_F(E_k^+)\right]\right] - \sum_k (\gamma_1-\gamma_2)\left[1-\left[n_F(E_k^-)-n_F(E_k^+)\right]\right] \bf{A} \nonumber\\
	&\xlongequal{T=0}&\sum_k(\gamma_1+\gamma_2)\left[1-\frac{\epsilon_k}{\mathcal{E}_k}\right] \bf{A} \nonumber\\
	&=& (\gamma_1+\gamma_2) n_c \bf{A}
\end{eqnarray}
where $n_c$ is the number of carriers in either band.

Below we shall calculate the optical conductance and check the optical sum rule
\begin{eqnarray}
	\Im \Pi_{\mu\nu}(\Omega,q\to 0) &=& -\frac{\pi}{V}\sum_{k}k_\mu^2\frac{(\gamma_1-\gamma_2)^2\Delta^2}{\mathcal{E}_k^2}\left[\delta(\Omega+2\mathcal{E}_k)-\delta(\Omega-2\mathcal{E}_k)\right] \delta_{\mu\nu} \nonumber\\
	&=& -\frac{\pi(\gamma_1-\gamma_2)^2 \epsilon_F N_{\gamma^+}(0)}{2(\gamma_1+\gamma_2)} \int d\epsilon  \frac{\Delta^2}{\epsilon^2+\Delta^2} \left[\delta\left(\Omega+2\sqrt{\epsilon^2+\Delta^2}\right)-\delta\left(\Omega-2\sqrt{\epsilon^2+\Delta^2}\right)\right]\delta_{\mu\nu}
\end{eqnarray}
Using the property of $\delta$-function $\delta\left[g(\epsilon)\right]=\sum_i\frac{\delta(\epsilon-\epsilon_i)}{\abs{g'(\epsilon_i)}}$, we have
\begin{eqnarray}
	\int d\epsilon  \frac{\Delta^2}{\epsilon^2+\Delta^2} \delta\left(\Omega-2\sqrt{\epsilon^2+\Delta^2}\right) &=& \int d\epsilon  \frac{\Delta^2}{\epsilon^2+\Delta^2} \left[\frac{\delta\left(\epsilon-\frac{1}{2}\sqrt{\Omega^2-4\Delta^2}\right)}{2\sqrt{\Omega^2-4\Delta^2}/\Omega} + \frac{\delta\left(\epsilon+\frac{1}{2}\sqrt{\Omega^2-4\Delta^2}\right)}{2\sqrt{\Omega^2-4\Delta^2}/\Omega}\right] \nonumber\\
	&=& \frac{4\Delta^2}{\Omega\sqrt{\Omega^2-4\Delta^2}},\quad (\Omega>0)
\end{eqnarray}
Since $\Im \Pi_{\mu\nu}(\Omega,q\to 0)$ is odd function of $\Omega$, the integration of it is
\begin{eqnarray}
	\int_{-\infty}^\infty d\Omega \frac{\Im \Pi_{\mu\nu}(\Omega,q\to 0)}{\Omega} &=& 2\int_0^\infty d\Omega \frac{\Im \Pi_{\mu\nu}(\Omega,q\to 0)}{\Omega} \nonumber\\
	&=& \frac{\pi(\gamma_1-\gamma_2)^2 \epsilon_F N_{\gamma^+}(0)}{\gamma_1+\gamma_2}\delta_{\mu\nu} \int_{2\Delta}^\infty d\Omega \frac{4\Delta^2}{\Omega^2\sqrt{\Omega^2-4\Delta^2}} \nonumber\\
	&=& \frac{\pi(\gamma_1-\gamma_2)^2 \epsilon_F N_{\gamma^+}(0)}{\gamma_1+\gamma_2}\delta_{\mu\nu} \nonumber\\
	&=& \pi\Re \Pi_{\mu\nu}(\Omega=0,q\to 0)
\end{eqnarray}
Thus the optical sum rule is satisfied.

The real part of optical conductance is
\begin{eqnarray}
	\Re \sigma_{\mu\mu}(q\to 0,\Omega>0) = \frac{\Im \Pi_{\mu\nu}(\Omega,q\to 0)}{\Omega} = \frac{4\pi(\gamma_1-\gamma_2)^2 \Delta^2\epsilon_F N_{\gamma^+}(0)}{2(\gamma_1+\gamma_2)\Omega^2\sqrt{\Omega^2-4\Delta^2}} 
\end{eqnarray}
which is showed in Fig.2(d) in the main text.

\section{Details on the numerical calculation}
In the main text, we numerically calculated the paramagnetic response contribution to the superfluid density of the lattice model. The Hamiltonian of the model under the basis $\Psi_{k}^\dagger=\{c_{1k\uparrow}^\dagger,c_{2k\uparrow}^\dagger,c_{1-k\downarrow},c_{2-k\downarrow}\}^\mathrm{T}$ can be written as
\begin{equation}
	H_{BdG}=\left(\begin{array}{cccc}
		\xi_{1k} & 0 & \Delta_{0} & \Delta_{1} \\
		0 & \xi_{2k} & \Delta_{1} & \Delta_{0} \\
		\Delta_{0} & \Delta_{1} & -\xi_{1-k} & 0 \\
		\Delta_{1} & \Delta_{0} & 0 & -\xi_{2-k}
	\end{array}\right), \label{lattice}
\end{equation}
where $\xi_{1,2}=-2t_{1,2}\left(\cos{k_x}+\cos{k_y}\right)-\mu$ is the lattice tight binding model including nearest neighbor hopping. $\Delta_0$ and $\Delta_1$ are intraband pairing and interband pairing order parameters respectively. 
Under this basis, the current operator can be expressed as 
\begin{equation}
	J^P_\mu=\left(\begin{array}{cccc}
		\partial\xi_{1k}/\partial k_\mu & 0 & 0 & 0 \\
		0 & \partial\xi_{2k}/\partial k_\mu & 0 & 0 \\
		0 & 0 & \partial\xi_{1-k}/\partial k_\mu & 0 \\
		0 & 0 & 0 & \partial\xi_{2-k}/ \partial k_\mu
	\end{array}\right). \label{current}
\end{equation}

The paramagnetic response can be calculated by the current-current correlated function
\begin{eqnarray}
	\Pi_{\mu\mu}(q,\ri \Omega_n)&=& \frac{1}{V\beta}\sum_{k,\ri\omega_n} \Tr[J^P_\mu(k,q)\mathcal{G}(k,\ri\omega_n)J^P_\mu(k,q)\mathcal{G}(k+q,\ri\omega_n+\ri\Omega_n)] \nonumber\\
	&=& \frac{1}{V\beta}\sum_{k,\ri\omega_n,o_{1-4},bb'} J^P_{\mu,o_1o_2}(k,q)U_{o_2b}(k)\mathcal{G}_{bb}(k,\ri\omega_n)U^*_{o_3b}(k)J^P_{\mu,o_3o_4}(k,q)U_{o_4b'}(k+q)\mathcal{G}_{b'b'}(k+q,\ri\omega_n+\ri\Omega_n)U^*_{o_1 b'}(k+q) \nonumber\\
	&=& \frac{1}{V}\sum_{k,bb'} \tilde{J}^P_{\mu,b'b}(k,q) \tilde{J}^{P\dagger}_{\mu, b'b}(k,q)\frac{n_F(E_k^b)-n_F(E_{k+q}^{b'})}{E_k^b-E_{k+q}^{b'}}.
	\label{para}
\end{eqnarray}
where the subscripts $o_1$ to $o_4$ represent the indices of the basis in the BdG Hamiltonian Eq.\ref{lattice}, and $b,b'$ denote the indices of the diagonalized BdG band basis. The Green's function in the band basis is $\mathcal{G}_{bb}=\frac{1}{\ri\omega_n-E_k^b}$. $E_k^b$ is the $b^\mathrm{th}$ eigenvalue of Eq.\ref{lattice} and $U(k)$ is the unitary matrix for diagonalization. $\tilde{J}^P_{\mu}(k,q)=U^\dagger(k+q)J^P_\mu(k,q)U(k)$ is the current operator under the band basis.

\begin{figure}
	\begin{center}
		\fig{4.0in}{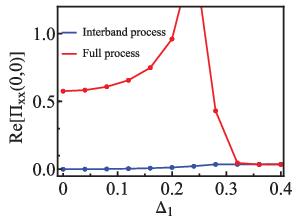}
		\caption{ The full paramagnetic response and the paramagnetic response considering only interband transition processes as functions of the interband pairing $\Delta_1$. The parameters used in the calculation are $t_1=1$, $t_2=0.6$, $\mu=1$, $T=0.01$.
			\label{full}}
	\end{center}
\end{figure}

In the calculation of the paramagnetic response shown in Fig.3(a) of the main text, we only considered the interband transition processes of the BdG bands, specifically the case where $b\neq b'$ in Eq.\ref{para}, and omitted the contributions from the Bogoliubov Fermi surface (BFS) when $b=b'$. This was done to more clearly demonstrate the finite paramagnetic response resulting from the asymmetric particle-hole interband transitions caused by interband pairing. In Fig.\ref{full}, we present the full paramagnetic response and the paramagnetic response considering only interband transition processes as functions of the interband pairing $\Delta_1$ when $t_1=1, t_2=0.6$. It can be observed that when $\Delta_1$ is small, the paramagnetic response is primarily contributed by the BFS. However, when $\Delta_1$ is large, the complete paramagnetic response converges with the saturation value of the response considering only interband transition processes. This indicates that even when the BFS vanishes, the system's superfluid density still exhibits a finite paramagnetic contribution due to interband pairing.

\section{The energy spectrum of different interband pairing}
\begin{figure}
	\begin{center}
		\fig{5.0in}{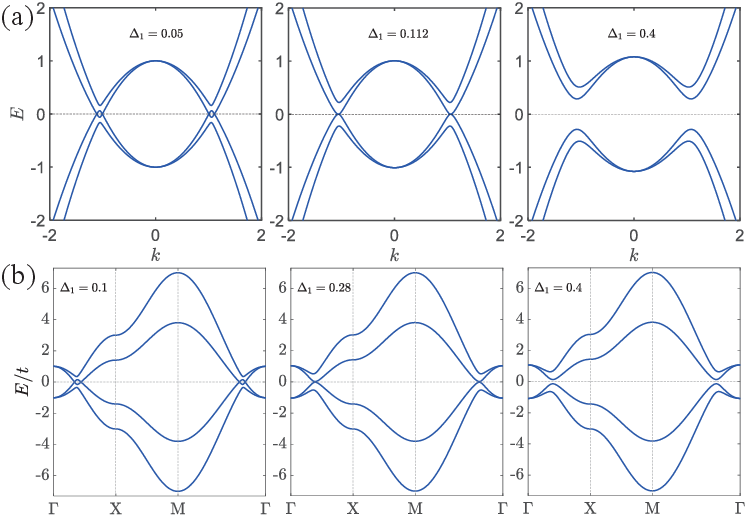}
		\caption{(a) The energy spectrum with different interband pairing in the continuum model. The parameters used in the calculation are $\gamma_1=1$, $\gamma_2=0.8$, $\mu=1$. (b) The energy spectrum with different interband pairing in the lattice model. The parameters used in the calculation are $t_1=1$, $t_2=0.6$, $\mu=1$.
			\label{figS2}}
	\end{center}
\end{figure}

Unlike intraband pairing, in our interband pairing model, the system is gapless at small $\Delta_1$ values due to the presence of a BFS. The energy spectrum only fully gapped when $\Delta_1$ exceeds a critical value $\Delta_c$. Fig.\ref{figS2} illustrates the gap opening process in both the continuum model and the lattice model. In the continuum model, the critical value is $\Delta_c=\frac{\mu\abs{\gamma_1-\gamma_2}}{2\sqrt{\gamma_1\gamma_2}}=0.112$. In the lattice model, the $\Delta_c$ is about 0.28, which is just the value of $\Delta_1$ at which the paramagnetic response reaches saturation in Fig.3(a) of the main text.

\section{Two different-mass bands with the same Fermi surface}

\begin{figure}
	\begin{center}
		\fig{5.0in}{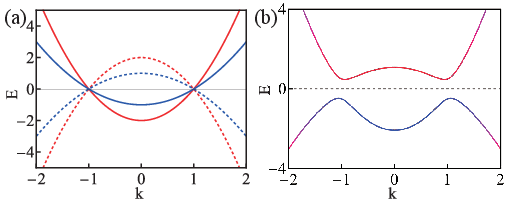}
		\caption{(a) The energy of the two bands with different mass and the same Fermi surface. The dashed line is the particle-hole symmetric band. (b) The band structure of the two different-mass band superconductor with inter-band pairing. The parameters used in the calculation are $\gamma_1=2$, $\gamma_2=1$, $\mu_1=2$, $\mu_2=1$ and $\Delta=0.5$.
			\label{figS3}}
	\end{center}
\end{figure}

In our main text, the two different-mass bands have a degeneracy at the band bottom, resulting in separated Fermi surfaces. Consequently, the interband pairing order parameter needs to exceed a critical value for the system to be fully gapped. Actually, the bottom of the two bands may not be degenerate, allowing their Fermi surfaces to coincide. That means the two bands hosts the dispersion with different Fermi energy $\xi_\alpha(k)=\frac{k^2}{2m_\alpha}-\mu_\alpha$. By tuning $\mu_1/\mu_2=\gamma_1/\gamma_2$, we can get this case as shown in Fig.\ref{figS3}(a). Due to the perfectly nesting Fermi surface of the two bands, infinitesimal interband pairing order parameter can make the system fully gapped as shown in Fig.\ref{figS3}(b).
We find that in this case, the eigenvalue of $H_1$ are $E_k^\pm=\eta_k\pm \sqrt{\epsilon_k^2+\Delta^2}$, where $\eta_k= \frac{\gamma_1-\gamma_2}{2}k^2-\frac{\mu_1-\mu_2}{2}$ and $\epsilon_k=\frac{\gamma_1+\gamma_2}{2}k^2-\frac{\mu_1+\mu_2}{2}$. It is a little different from the case in the main text, but the result of superfuild response is the same.

\section{Quasiparticle band of PDW}
\begin{figure}
	\begin{center}
		\fig{5in}{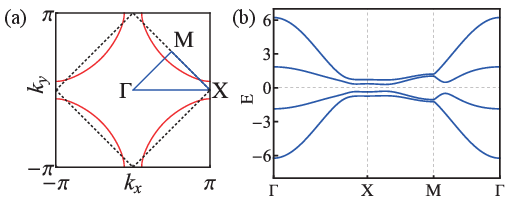}
		\caption{(a) The first Brillouin zone (BZ) of the square lattice. Red lines show the Fermi surface of $\xi_k$. Dashed lines are the reduced BZ. Blue lines are the high symmetry line.  (b) The quasiparticle band of the Hamiltonian Eq.6 in the main text along the high symmetry line. The order parameter is set as $\Delta_{on}=0.5$ and $\Delta_{PDW}=0.2$.
			\label{figS4}}
	\end{center}
\end{figure}

We plot the quasiparticle band of the $Q=(\pi,\pi)$ PDW in Fig.\ref{figS4}(b) along the high symmetric line shown in Fig.\ref{figS4}(a). The optical transition process is similar to that of the interband pairing SC with different mass Cooper pairs.

\section{Superfluid response and optical conductance of the charge 4$e$ SC}
Following the paper \cite{charge4e}, the current operator matrix under the basis $\left(c_{1 \bm{k} \uparrow},\;c_{2 \bm{k} \uparrow},\;c_{1-\bm{k} \downarrow}^{\dagger},\;c_{2-\bm{k} \downarrow}^{\dagger}\right)^\mathrm{T}$ can be written as 
\begin{eqnarray}
	J^P(k,q)=\frac{2k+q}{2m} \tau_0\sigma_0
\end{eqnarray}
where $\tau$ and $\sigma$ are Pauli matrix in particle-hole and orbital space respectively.
The single particle Green's function has the form 
\begin{equation}
	\bar{G}^R(\bm{k},\ri\omega_n) = \left(\begin{array}{cccc}
		\frac{\omega+\ri\eta+3\xi_\bk}{(\omega+\ri\eta+\xi_\bk)^2-E_\bk^2} & & & \\
		& \frac{\omega+\ri\eta+3\xi_\bk}{(\omega+\ri\eta+\xi_\bk)^2-E_\bk^2} & & \\
		& & \frac{\omega+\ri\eta-3\xi_\bk}{(\omega+\ri\eta-\xi_\bk)^2-E_\bk^2} & \\
		& & & \frac{\omega+\ri\eta-3\xi_\bk}{(\omega+\ri\eta-\xi_\bk)^2-E_\bk^2}
	\end{array}\right).
\end{equation}
Then the current-current correlation function can be calculated via Eq.\ref{JJ} and the result at zero temperature are
\begin{equation}
	\Re \Pi_{\mu\mu}(\ri\Omega_n=0,q\to 0) = \frac{\epsilon_F N(\epsilon_F)}{6m}
\end{equation}
\begin{equation}
	\Im \Pi_{\mu\mu}(\Omega,q\to 0)  = \frac{\pi\epsilon_F N(\epsilon_F)}{3m}\frac{\Delta^2}{\Omega\sqrt{\Omega^2-4\Delta^2}}
\end{equation}
Thus, the superfluid density is obtained as
\begin{equation}
	\rho_s = \la \frac{n}{m}\ra - \Re \Pi_{\mu\mu}(\ri\Omega_n=0,q\to 0) = \frac{3\rho}{4}
\end{equation}
where $\rho=\frac{2\epsilon_F N(\epsilon_F)}{3m}$,
and the real part of the optical conductance is 
\begin{eqnarray}
	\Re \sigma_{\mu\mu}(q\to 0,\Omega>0) = \frac{\Im \Pi_{\mu\nu}(\Omega,q\to 0)}{\Omega} = \frac{\pi\epsilon_F N(\epsilon_F)}{3m}\frac{\Delta^2}{\Omega^2\sqrt{\Omega^2-4\Delta^2}}
\end{eqnarray}
It's straightforward to check that the optical sum rule is satisfied.


\end{document}